\def\year{2020}\relax
\documentclass[letterpaper]{article} 
\usepackage{aaai20}  
\usepackage{times}  
\usepackage{helvet} 
\usepackage{courier}  
\usepackage[hyphens]{url}  
\usepackage{graphicx} 
\usepackage{graphicx}  
\usepackage{booktabs}

\usepackage{amsmath,amssymb,amsfonts}
\usepackage{graphicx}
\usepackage{textcomp}
\usepackage{times}
\usepackage{xcolor}
\usepackage{soul}

\usepackage{subfigure}

\usepackage{algorithm}
\usepackage{algpseudocode}
\usepackage{mathtools}

\usepackage{times}
\usepackage{soul}
\usepackage{graphicx}
\usepackage{amsmath}
\usepackage{comment}
\usepackage{mathtools}
\usepackage{amsthm}
\newtheorem{proposition}{Proposition}

\def\BibTeX{{\rm B\kern-.05em{\sc i\kern-.025em b}\kern-.08em
    T\kern-.1667em\lower.7ex\hbox{E}\kern-.125emX}}

\begin{document}
\title{COBRA: Context-aware Bernoulli Neural Networks for Reputation Assessment }
\author{\Large \textbf{Leonit Zeynalvand,\textsuperscript{\rm 1} Tie Luo,\textsuperscript{\rm 2} Jie Zhang\textsuperscript{\rm 1}}\\ 
\textsuperscript{\rm 1}School of Computer Science and Engineering, Nanyang Technological University, Singapore
\\ 
\textsuperscript{\rm 2}Department of Computer Science, Missouri University of Science and Technology, USA\\
leonit001@e.ntu.edu.sg, tluo@mst.edu, zhangj@ntu.edu.sg 
}
\maketitle

\begin{abstract}
Trust and reputation management (TRM) plays an increasingly important role in large-scale online environments such as multi-agent systems (MAS) and the Internet of Things (IoT). One main objective of TRM is to achieve accurate trust assessment of entities such as agents or IoT service providers. However, this encouters an {\em accuracy-privacy dilemma} as we identify in this paper, and we propose a framework called {\em \underline{Co}ntext-aware \underline{B}ernoulli Neural Network based \underline{R}eputation \underline{A}ssessment} (COBRA) to address this challenge. 
COBRA encapsulates agent interactions or transactions, which are prone to privacy leak, in machine learning models, and aggregates multiple such models using a Bernoulli neural network to predict a trust score for an agent. COBRA preserves agent privacy and retains interaction contexts via the machine learning models, and achieves more accurate trust prediction than a fully-connected neural network alternative. COBRA is also robust to security attacks by agents who inject fake machine learning models; notably, it is resistant to the 51-percent attack. The performance of COBRA is validated by our experiments using a real dataset, and by our simulations, where we also show that COBRA outperforms other state-of-the-art TRM systems.
\end{abstract}

\section{Introduction} \label{intro}
Trust and reputation management (TRM) systems are critical to large-scale online environments such as multi-agent systems (MAS) and the Internet of Things (IoT), where agents\footnote{Throughout this paper, we use the term agents in a broader sense which is not limited to agents in MAS, but also includes IoT service providers and consumers as well as other similar cases.} are more autonomous and tend to have more interactions with each other. Without a reliable TRM system, such interactions will be significantly hindered due to lack of trust in services or information provided by other agents. Formal contracts such as Service Level Agreements (SLA) are hard to enforce in such open environments because of high cost and lack of central authorities.

Early TRM systems such as \cite{josang2002beta} rely on \textit{first-hand evidence} to derive trust scores of agents. For example, an agent \textit{Alice} assigns a trust score to another agent \textit{Bob} based on the outcome of her previous interactions with \textit{Bob}. However, as the scale of the systems grows (e.g., IoT), first-hand evidence becomes too sparse to support reliable trust evaluation. Hence, \textit{second-hand evidence} was exploited by researchers to supplement first-hand evidence. In that case, \textit{Alice} would assign a trust score to \textit{Bob} based not only on her own interactions with \textit{Bob} but also on what other agents advise about \textit{Bob}. 

However, what form the second-hand evidence should take has been largely overlooked. This engenders an important issue which we refer to as the \textit{accuracy-privacy dilemma}. To illustrate this, suppose {\it Alice} consults another agent \textit{Judy} about how trustworthy \textit{Bob} is. One way is to let \textit{Judy} give a trust score or rating about \textit{Bob} \cite{yu2013survey}, which is the approach commonly adopted in the trust research community. This approach is simple but loses the {\em context information} of the interactions between agents. For example, the context could be the transaction time and location, and service provided by an agent during off-peak hours could have higher quality (more SLA-conformant) than during peak hours. Without such context information, trust assessment based on just ratings or scores would have lower accuracy. 
On the other hand, another method is to let \textit{Judy} reveal her entire interaction history with \textit{Bob} (e.g., in the form of a detailed review), 
which is the approach commonly used in recommender systems. Although the information disclosed as such is helpful for trust assessment of {\it Bob}, it is likely to expose substantial privacy of {\it Bob} and {\it Judy} to {\it Alice} and possibly the public.\footnote{Recommender systems can take this approach because they are generally considered trusted intermediaries, and they focus on preference modeling rather than trust and reputation modelling.}

To address this accuracy-privacy dilemma, and in the meantime avoid relying on a trusted third-party which is often not available in practice, we propose a framework called {\em \underline{Co}ntext-aware \underline{B}ernoulli Neural Network based \underline{R}eputation \underline{A}ssessment} (COBRA). It encapsulates the detailed second-hand evidence using machine learning models, and then aggregate these model using a Bernoulli neural network (BNN) to predict the trustworthiness of an agent of interest (e.g., an IoT service provider). The encapsulation protects agent privacy and retains the context information to enable more accurate trust assessment, and the BNN accepts the outputs of those ML models and the information-seeking agent's ({\it Alice'} as in the above example) first-hand evidence as input, to make more accurate trustworthiness prediction (of {\it Bob} as in the above example). 

The contributions of this paper are summarized below:

\begin{itemize}
  
  \item We identify the accuracy-privacy dilemma and propose COBRA to solve this problem using a model encapsulation technique and a Bernoulli neural network. COBRA preserves privacy by encapsulating second-hand evidence using ML models, and makes accurate trust predictions using BNN which fuses both first-hand and second-hand evidence, where the valuable context information was preserved by the ML models.

  \item The proposed BNN yields more accurate predictions than the standard fully-connected feed-forward neutral networks, and trains significantly faster. In addition, it is also general enough to be applied to similar tasks when the input is a set of probabilities associated with Bernoulli random variables. 
 
  \item The design of COBRA takes security into consideration and it is robust to fake ML models; in particular, it is resistant to the 51-percent attack, where the majority of the models are compromised.

  \item We evaluate the performance of COBRA using both experiments based on a real dataset, and simulations. The results validate the above performance claims and also show that COBRA outperforms other state-of-the-art TRM systems.
\end{itemize}

\section{Related Work}
A large strand of literature has attempted to address TRM in multi-agent systems. The earliest line of research had a focus on first-hand evidence ~\cite{yu2013survey}, using it as the main source of trustworthiness calculation. For example, Beta reputation system ~\cite{josang2002beta} proposes a formula to aggregate first-hand evidence represented by binary values indicating positive or negative outcomes of interaction. Concurrently with the spike in popularity of recommender systems in late 2004 ~\cite{recSys,fang2015multi}, the alternative usage of TRM in preference and rating management gained much research attention. However, the binary nature of trust definition presents a barrier because recommender systems conventionally use non-binary numerical ratings. To this end, Dirichlet reputation systems ~\cite{josang2007dirichlet} generalize the binomial nature of beta reputation systems to accommodate multinomial values.

A different line of research focuses on second-hand evidence~\cite{yu2013survey} as a supplementary source of trustworthiness calculation. These works calculate a trust score by aggregating second-hand evidence and a separate trust score by aggregating first-hand evidence, and then a final score by aggregating these two scores. Some early trust models such as ~\cite{josang2002beta} are also applicable to second-hand evidence. The challenges in this line of research are \cite{zhang2008evaluating}: (i) How to determine which second-hand evidence is less reliable, since second-hand evidence is provided by other agents? (ii) How much to rely on trust scores that are derived from second-hand evidence compared to scores derived from first-hand evidence? 

To address the first challenge, the Regret model~\cite{sabater2001regret} assumes the existence of social relationships among agents (and owners of agents), and assigns weights to second-hand evidence based on the type and the closeness of these social relationships. These weights are then used in the aggregation of second-hand evidence. More sophisticated approaches like Blade~\cite{blade} and Habit ~\cite{habit} tackle this issue with a statistical approach using Bayesian networks and hence do not rely on heuristics. To address the second challenge, \cite{fullam2007dynamically} uses a Q-learning technique to calculate a weight which  determines the extent to which the score derived from second-hand evidence affects the final trust score.

A separate thread of research relies solely on stereotypical intrinsic properties of the agents and the environment in which they operate, to derive a likelihood of trustworthiness without using any evidence. These approaches ~\cite{sc1,sc2,sc3,sc4} are considered a complement to evidence-based trust and are beneficial when there is no enough evidence available.

Our proposed approach does not fall under any of these categories; instead, we introduce model encapsulation as a new way of incorporating evidence into TRM. We make no assumptions on the existence of stereotypical or socio-cognitive information, as opposed to  ~\cite{sc1,sc2,sc3,sc4,sabater2001regret}). Our approach has minimal privacy exposure, which is unlike ~\cite{blade,habit}, and preserves important context information.

\section{Model Encapsulation} \label{modeling}


COBRA encapsulates second-hand evidence in ML models, which achieves two purposes: (i) it preserves privacy of agents who are involved in the past interactions; (ii) it retains context information which will help more accurate trust prediction later (described in the next section). 


In this technique, each agent trains a ML model using its past interaction records with other agents in different contexts. Specifically, an agent $u \in \mathcal{A}$ (the set of all the agents) trains a model $\mathcal{M}_u^z(\zeta)=p$ based on its past direct interaction (i.e., first-hand evidence) with an agent $z$. The input to the model is a set $\mathcal{\zeta}$ of context features (e.g., date, time, location), and the output is a predicted conditional probability $p$ indicating how trustworthy $z$ is for a given context $\mathcal{\zeta}$.

To build this model, the agent $u$ maintains a dataset that records its past interactions with each other agent, where each record includes the context $\zeta$ and the outcome $t \in \mathbb{Z}_2=\{0,1\}$ with 0 and 1 meaning a negative and a positive outcome respectively (e.g., whether the SLA is met or not). For non-binary outcomes, they can be handled by using the common method of converting a multi-class classification problem into multiple binary classification problems (so there will be multiple models for each agent).
Then, agent $u$ trains a machine learning model for each agent, say $z$, using the corresponding dataset to obtain $\mathcal{M}_u^z(\zeta)=p$.

COBRA does not restrict the choice of ML models and this is up to the application and the agents. For example, agents hosted on mobile phones can choose simple models such as decision trees and Naive Bayes, while those on desktop computers or in the cloud can use more complex models such as random forests and AdaBoost. Furthermore, agents can choose different models in the same application, meaning that $\mathcal{M}_{u_1}^z$ may not be the same type as $\mathcal{M}_{u_2}^z$. On the other hand, the context feature set $\zeta$ needs to be fixed for the same application.

\begin{figure}[t]
\centering\includegraphics[width=1.04\linewidth]{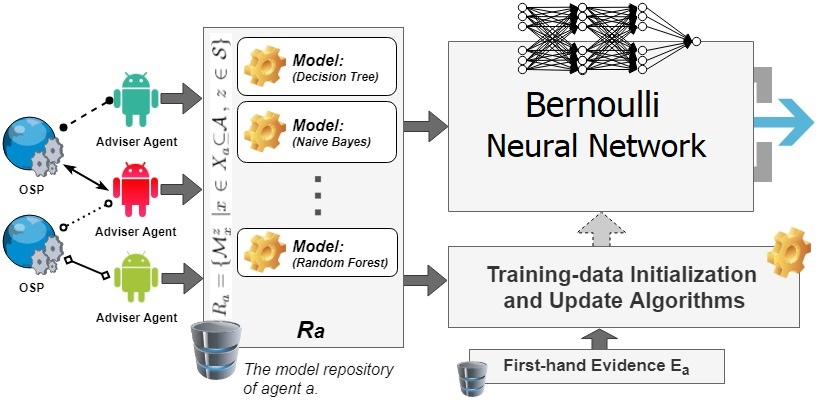}
\caption{The COBRA framework.}
\label{fig:diagram}
\end{figure}

{\em Model Sharing.}
Whenever an agent, say $a$, seeks advice (i.e., second-hand evidence) from another agent, say $u$, about an agent of interest, say $z$, the agent $u$ can share its model $\mathcal{M}_u^z(\zeta)$ with $a$. This avoids exposing second-hand evidence directly and thereby preserves privacy of both $u$ and $z$. It also retains context information as compared to $u$ providing $a$ with just a single trust score of $z$, and hence helps more informed decision making in the subsequent step (described in Section \ref{deep}).

Note that the information we seek to keep private is the contextual details of the interactions between $u$ and $z$, whereas concealing the identities of $u$ and $z$ is not the focus of this work. 

Sharing a model is as straightforward as transferring the model parameters to the soliciting agent (i.e., $a$ in the above example), or making it accessible to all the agents (in a read-only mode). This sharing process does not require a trusted intermediary because the model does not present a risk of privacy leaking about $u$ and $z$. The required storage is also very low as compared to storing the original evidence.

Moreover, COBRA does not assume that all or most models are accurate. Unlike many existing work assuming \textit{honest majority} and hence being vulnerable to 51-percent attack, COBRA use a novel neural network architecture (Section \ref{deep}) that is more robust to model inaccuracy and even malice (e.g., models that give opposite outputs). 

\section{Bernoulli Neural Network} \label{deep}

After model encapsulation which allows for a compressed transfer of context-aware second-hand evidence with privacy preservation, the next question is how to aggregate these models to achieve an accurate prediction of the trustworthiness of a target agent. Using common measures of central tendency such as mean, mode, etc. will yield misleading results because an adviser agent's ($u$'s) model was trained on a dataset with likely different contexts than the advisee agent's ($a$'s) context. In a sense, this problem is akin to the problem found in {\em transfer learning}. Besides, COBRA aims to relax the assumption of \textit{honest majority} and give accurate predictions even when the majority of the models are inaccurate or malicious.

In this section, we propose a solution based on artificial neural networks (ANN). The reasons for choosing ANN are two. First, the task of predicting trustworthiness in a specific context given other agents' models, is a linearly non-separable task with high dimensional input space (detailed in Section \ref{deepdown1}). Such tasks can specifically benefit from ANN's capability of discovering intrinsic relationships hidden in data samples~\cite{zhang2010computational}. Second, the models are non-ideal due to the possibly noisy agent datasets, but ANN is highly noise tolerant and sometimes can even be positively affected by noise~\cite{madic2010assessing,luo2014deep}.

Therefore, we propose a Bernoulli Neural Network (BNN) as our solution. BNN specializes in processing data that is a set of probabilities associated with random variables of Bernoulli distribution, which perfectly matches our input space which is a set of predicted trust scores between zero and one indicating the probability of an agent being trustworthy in a given context. In contrast to the widely used Convolutional Neural Network (CNN), BNN does not require data to have a grid-like or structured topology, and hence matches well with trust or reputation scores. Specifically, unlike CNN which uses the hierarchical pattern in data, BNN uses information entropy, to determine the connections in the network. 

Fig. \ref{fig:diagram} provides an overview of COBRA, where the models on the left hand side are from the encapsulation technique described in Section \ref{modeling}, and the right hand side is the BNN described in this section. In the following, we explain the architectural design of BNN in Section \ref{deepdown1}-3 and assemble the data required for training the BNN in Section \ref{deepdown4}. 

\subsection{Topology} \label{deepdown1}
We propose a $(N+1)$-layer network architecture for the BNN, where the input layer is denoted by $L_0$, the output layer by $L_N$, and the hidden layers by $L_k$ where $k=1,...,N-1$. The weight of an edge $(i,j)$ is denoted as $w_{ij}$ where $i \in L_{k-1}$ and $j \in L_{k}$, $k=1,2,...,N$. The bias at layer $L_{k-1}$ is $b_{k-1}$. Thus, the entire network can be compiled from Eq. \ref{deq:1} where the output of any node $j\in L_k$ is given by $y_j$. The inputs $x_j$ of the network in Eq. \ref{deq:1} are assembled from (1) the models explained in Section \ref{modeling} and (2) the context features, where the assembling process is explained in Section \ref{deepdown4}. The former are values between zero and one which indicate the predicted trustworthiness probability of an entity in the given context sourced from different predictors. 

\begin{figure*}[t]

\begin{align}\label{deq:1}
  y_j = \begin{cases}
      f_k\left({\displaystyle b_{k-1} + \sum_{i\in L_{k-1}} sgn\left({\displaystyle\left\lfloor{}\dfrac{|L_k|-h(y_i)|L_k|}{\left|\left\{ y_{t} | \ \  \dfrac{\partial y_t}{\partial y_i} \neq 0 \, , \,  t \in L_{k}\right\}\right|+1}\right\rfloor}\right)w_{ij} y_i}\right), & j \in L_{k} : k=1,2,...,N \\
      x_j, & j \in L_{0}
    \end{cases}
\end{align}

\end{figure*}

The probabilistic nature of the inputs, enables us to calculate how informative an input is, by calculating the entropy of the predicted trustworthiness for which the input indicates the probability. This is used in Eq. \ref{deq:1} to ensure that the number of neural units an input connects to is inversely proportional to the average of information entropy calculated for input samples. Each sample of an input in the training data-set (exclusive of context features) is a probability (sourced prediction) associated with a random variable with Bernoulli distribution (trustworthiness). Hence, $h(.)$ is defined recursively by Eq. \ref{eq:h} where $\overline{H}{(x_j)}$ is the average entropy of $x_j$ in the training data-set. For context features $\overline{H}{(x_j)}=0$ because the values of these features are not probabilities of Bernoulli random variables and hence the notion of entropy can only be applied to their entire feature space not individual values which is what is used in Eq. \ref{deq:1}. 

Moreover, $f_k(.)$ in Eq. \ref{deq:1} is the activation function of layer $k$. The design choice of the activation functions is explained in Section \ref{deepdown3}.

\begin{align}
  { h(y_j) =}
    \begin{cases}\displaystyle
      \dfrac{\displaystyle\sum_{i \in I}{h(y_{i})}}{|I|} , & j\in L_{k} \, : \, I=\{i \in L_{k-1} | \dfrac{\partial y_j}{\partial y_{i}} \neq 0\}  \\
         &  \\
      {\displaystyle \overline{H}{(x_j)}}, & j\in L_{0} 
    \end{cases} \label{eq:h}      
\end{align}

\subsection{Depth and width} \label{deepdown2}

The depth of BNN is $N$ since the input layer is not counted by convention. A feed-forward network with two hidden layers can be trained to represent any arbitrary function, given sufficient width and density (number of edges)~\cite{heaton2008introduction}. Our goal is to find the function which most accurately weights the predictions sourced from multiple predictors (i.e. high dimensional input space). Many of such sources can be unreliable or misleading either unintentionally (e.g. malfunction) or deliberately (e.g. malicious). There often does not exist a single source which is always reliable and some sources are more reliable in some contexts. Moreover the malicious sources sometimes collude with each other to make the attack harder to detect. Therefore, the function that we aim to estimate in this linearly non-separable task can have any arbitrary shape. Hence, we choose $N=3$ in our design to benefit from two hidden layers which suffice to estimate the aforementioned function as we demonstrate by our experiment results in Section \ref{evSec}.

The width of a layer is the number of units (nodes) in that layer, and accordingly we denote the width of a layer $k$ by $\vert L_k \vert$. Determining the width is largely an empirical task, and there are many rule-of-thumb methods used by the practitioners. For instance, \cite{heaton2008introduction} suggests that the width of a hidden layer be $2/3$ the width of the previous layer plus the width of the next layer. Inspired by this method we propose a measure called \textit{output gain} defined as the summation of the information gain of the inputs of a node  and determine $|L_k|$ by Eq. \ref{deq:2}. The width $|L_N|$ is set to $1$ because the network has only a single output which is the trust score (probability of being trustworthy). And the width $|L_0|$ is set to the total number of input nodes denoted by $n$. 

\begin{align}
  {\vert L_k \vert=}
    \begin{cases}\displaystyle
    n, & k=0 \\
      & \\
      {\displaystyle\left\lceil\sum_{j\in L_{0}} {\frac{2}{3}\left( 1 - \overline{H}(x_j) \right)}\right\rceil} + \vert L_{2} \vert , & k=1 \\
         & \\
        {\displaystyle\left\lceil {\frac{2}{3} \vert L_{1} \vert }\right\rceil} + \vert L_{3} \vert , & k=2 \\
         & \\
      1, & k=3
    \end{cases} \label{deq:2}      
\end{align}


\subsection{Activation and loss functions} \label{deepdown3}
Let us recall the activation function $f_k(.)$ from Eq. \ref{deq:1} in Section \ref{deepdown1}. Since we choose $N=3$ as explained in Section \ref{deepdown2}, we need to specify three activation functions $f_{1}(.), f_{2}(.)$, and $f_{3}(.)$ for the first hidden layer, second hidden layer, and the output layer, respectively.

For the output layer, we choose the sigmoid logistic function $f_3(z)=1/(1+e^{-z})$ because we aim to output a trust score (the probability that the outcome of interacting with a certain agent is positive for a given context). For the hidden layers, we choose the rectified linear unit (ReLU)~\cite{lecun2015deep} function as $f_{1,2}(z)=max(0,z)$, because the focus of hidden layers is to exploit the compositional hierarchy of the inputs to compose higher level (combinatoric) features so that data become more separable in the next layer, and hence the speed of convergence is a main consideration.

The weights in the BNN are computed using \textit{gradient descent back propagation} during the training process. However, sigmoid activation functions, as we choose, have a saturation effect which will result in small gradients, while gradients need to be large enough for weight updates to be propagated back into all the layers. Hence, we use 
{\em cross-entropy} $H(p,q) = - \sum_{x} p(x) \log(q(x))$ as the loss function to mitigate the saturation effect of the sigmoid function. Specifically, the $\log(.)$ component in the loss function counteracts the $\exp(.)$ component in the sigmoid activation function, thereby enabling gradient-based learning process.

\subsection{{Assembling training data }} \label{deepdown4}

Having explained the architectural design aspects of our Bernoulli neural network, now we explain its computational aspects. 

The output of the neural network is a predicted probability that a target agent $z$ is trustworthy (e.g., meets SLA) in a certain context $\zeta$, which (the probability) is what an agent $a$ tries to find out. The input of the network consists of (1) all the context features $\varsigma \in \zeta$ and (2) all the probabilities predicted by models $\mathcal{M}_u^z(\zeta)$ shared by all the $u\in X_a$ where $X_a \subseteq \mathcal{A}$ is the agents $a$ is seeking advice from. In the case that some agents $u \in X_a$ do not share their models with agent $a$, the corresponding input probability will be set to 0.5 to represent absolute no information. 
Formally, the input from the models to the neural network is given by
\begin{equation}
  G_a(x,z,\zeta) =
    \begin{cases}
        \mathcal M_{x}^{z}(\zeta) & \mathcal M_{x}^{z} \in R_a \\
        0.5 & M_{x}^{z} \not\in R_a
    \end{cases} \label{deq:6}     
\end{equation}
where $R_a$ is the set of models available to $a$. Most precisely, each input variable (to layer $L_0$) is specified by
\begin{equation}
  x_j = \begin{cases}
         \varsigma_{j}   & j=1,2,...,|\zeta| \\
         G_a(X_a^{j-|\zeta|},z,\zeta)  & j=|\zeta|+1,...,|\zeta|+|X_a|
    \end{cases} \label{deq:6b} 
\end{equation}
which also gives the number of input nodes (i.e. input dimension)
\[ n=|X_a|+|\zeta|. \]

\begin{algorithm}[t]
\footnotesize
  \caption{Training data initialization}\label{alg1}
  \textbf{Input:} First-hand evidence of agent $a$:\\
  $E_a=\{(z,\varsigma_1,\varsigma_2,...,\varsigma_{k},t)\ \  \vert z \in \mathcal{S}_a \subseteq \mathcal{S},  k=|\zeta|,  t \in \{0,1\}\}$ where $S$ is the set of target agents (whose reputation is to be predicted), 
  $R_a= \{\mathcal{M}_u^z \ \ \vert x \, \in \,   X_a{\subseteq}\mathcal{A} \ , \  z \, \in \, \mathcal{S}\}$

  \textbf{Output:} Training dataset ($Features$ and $Label$)

  \begin{algorithmic}[1]
        \State $Features, Label, temp \xleftarrow[\text{}]{\text{set}} \emptyset$ 
        \State $i,j \xleftarrow[\text{}]{\text{set}} 0$
        \For{{ each $(z,\varsigma_1,\varsigma_2,...,\varsigma_{k},t)$ in $E_a$}}
                \For{{each $u$ in $X_a$}}
                    \State $tmp[i] \xleftarrow[\text{}]{\text{set}} G_a(x,z,\varsigma_1,\varsigma_2,...,\varsigma_{k})$
                    \State $i \xleftarrow[\text{}]{\text{set}} i+1$
                \EndFor
                \State $Features[j] \xleftarrow[\text{}]{\text{set}} (\varsigma_1,\varsigma_2,...,\varsigma_{k},tmp[0],...,tmp[|X_a|])$
                \State $Label[j] \xleftarrow[\text{}]{\text{set}} t$
                \State $j \xleftarrow[\text{}]{\text{set}} j+1$
        \EndFor
      \State \textbf{return} $Features, Label$ 
  \end{algorithmic}
\end{algorithm}
\begin{algorithm}[t]
\footnotesize
  \caption{Update training data vertically}\label{alg2}
  \textbf{Input:} A new first-hand evidence $(z,\varsigma_1,\varsigma_2,...,\varsigma_{k},t)\ $ where $ z \in \mathcal{S}_a \subseteq \mathcal{S},  k=|\zeta|,  t \in \{0,1\}\}$\\ Current training dataset: $Features$ and $Label$\\
  Model repository: $R_a= \{\mathcal{M}_u^z \ \ \vert x \, \in \,   X_a{\subseteq}\mathcal{A} \ , \  z \, \in \, \mathcal{S}\}$.

  \textbf{Output:} Updated training data-set (new $Features$ and $Label$).

  \begin{algorithmic}[1]
        \State $i \xleftarrow[\text{}]{\text{set}} 0$
                \For{{each $u$ in $X_a$}}
                    \State $tmp[i] \xleftarrow[\text{}]{\text{set}} G_a(x,z,\varsigma_1,\varsigma_2,...,\varsigma_{k})$
                    \State $i \xleftarrow[\text{}]{\text{set}} i+1$
                \EndFor
                \State $Features[j] \xleftarrow[\text{}]{\text{append}} (\varsigma_1,\varsigma_2,...,\varsigma_{k},tmp[0],...,tmp[|X_a|])$
                \State $Label[j] \xleftarrow[\text{}]{\text{append}} t$
      \State \textbf{return} $Features, Label$ 
  \end{algorithmic}
\end{algorithm}

\begin{algorithm}[t]
\footnotesize
  \caption{Update training data horizontally}\label{alg3}
  \textbf{Input:} A new model $\mathcal{M}_{u'}^{z'}\ $ where $u' \, \in \, \mathcal{A}$ and $ z' \in  \mathcal{S}$\\ Current training data-set: $Features$ (no need for $Label$)\\First-hand evidence of agent $a$:\\
  $E_a=\{(z,\varsigma_1,\varsigma_2,...,\varsigma_{k},t)\ \  \vert z \in \mathcal{S}_a \subseteq \mathcal{S},  k=|\zeta|,  t \in \{0,1\}\}$\\
  \textbf{Output:} Updated training data-set (new $Features$).

  \begin{algorithmic}[1]
        \State $i \xleftarrow[\text{}]{\text{set}} 0$
        \For{{ each $(z,\varsigma_1,\varsigma_2,...,\varsigma_{k},t)$ in $E_a$}}
                \If{$z=z'$} 
                    \State $Features[i].u' \xleftarrow[\text{}]{\text{set}} \mathcal{M}_{u'}^{z'}(\varsigma_1,\varsigma_2,...,\varsigma_{k})$ 
                \EndIf
                \State $i \xleftarrow[\text{}]{\text{set}} i+1$
        \EndFor

      \State \textbf{return} $Features$ 
  \end{algorithmic}
\end{algorithm}

Thus, we transform the recursive Eq. \ref{deq:2} into a system of linear equations:
\begin{align}
    \begin{cases}
      |L_1|&=\frac{2}{3}\displaystyle\left(\left\lceil\sum_{j\in L_0} {\left( 1 - \overline{H}(x_j) \right)}\right\rceil\right)+|L_2| \\ \\
      |L_2|&=\frac{2}{3}(|L_1|)+1
    \end{cases} \label{deq:7}      
\end{align}
Solving Eq. \ref{deq:7} yields the widths of all the layers of our neural network: 
\begin{align}
    \begin{split}
    |L_1|&=2\displaystyle\left(\left\lceil\sum_{j\in L_0} {\left( 1 - \overline{H}(x_j) \right)}\right\rceil\right)+3,\\ |L_2|&=\left\lfloor{\frac{4}{3}\displaystyle\left(\left\lceil\sum_{j\in L_0} {\left( 1 - \overline{H}(x_j) \right)}\right\rceil\right)}\right\rfloor+3,\\
    |L_3|&=1.
\end{split}
\end{align} 

The weights are calculated using gradient descent back propagation based on training data. The training data is initialized once using Algorithm \ref{alg1} and updated \textit{vertically} upon acquiring new first-hand evidence using Algorithm \ref{alg2} and updated \textit{horizontally} upon acquiring a new model using Algorithm \ref{alg3}.

\begin{figure*}[t]
    \centering 
\subfigure[]{\label{fig:a}\includegraphics[width=58mm]{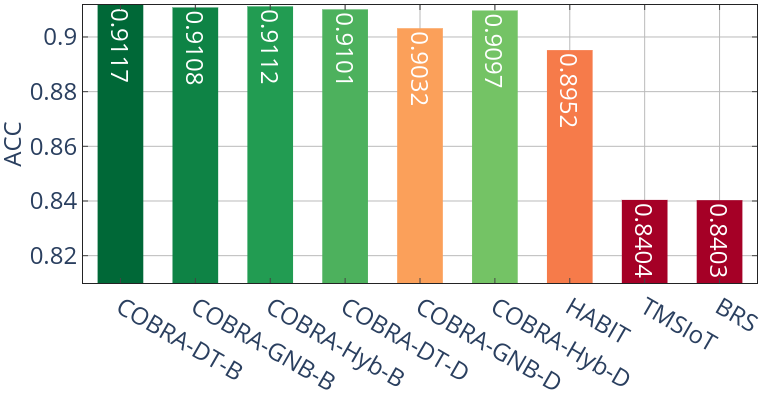}}
\renewcommand{\thesubfigure}{(c)}
\subfigure[]{\label{fig:c}\includegraphics[width=60mm]{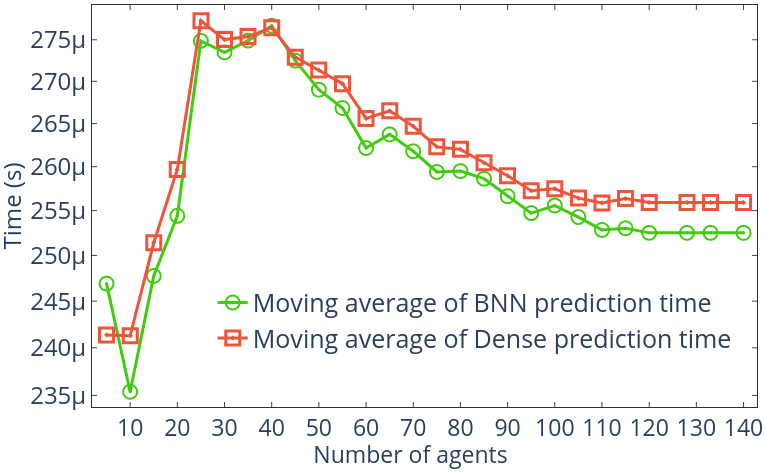}}
\renewcommand{\thesubfigure}{(e)}
\subfigure[]{\label{fig:e}\includegraphics[width=58mm]{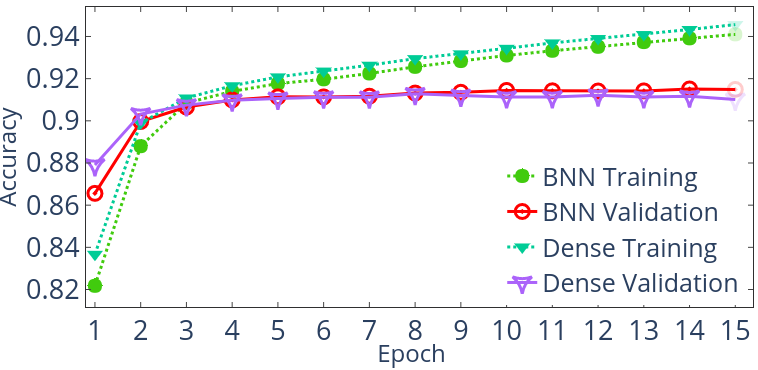}}

\medskip
\renewcommand{\thesubfigure}{(b)}
\subfigure[]{\label{fig:b}\includegraphics[width=58mm]{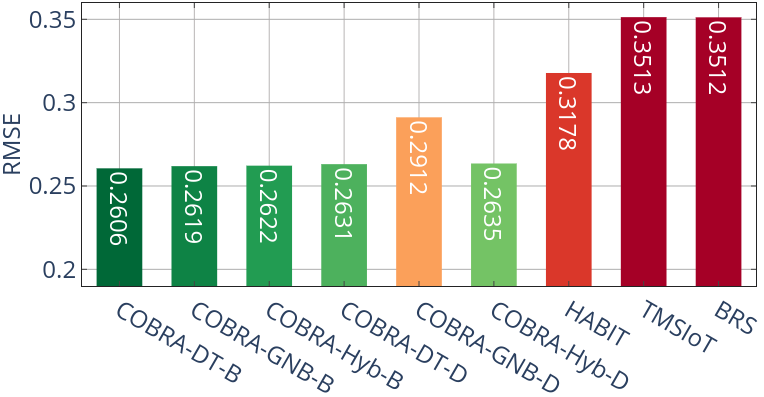}}
\renewcommand{\thesubfigure}{(d)}
\subfigure[]{\label{fig:d}\includegraphics[width=60mm]{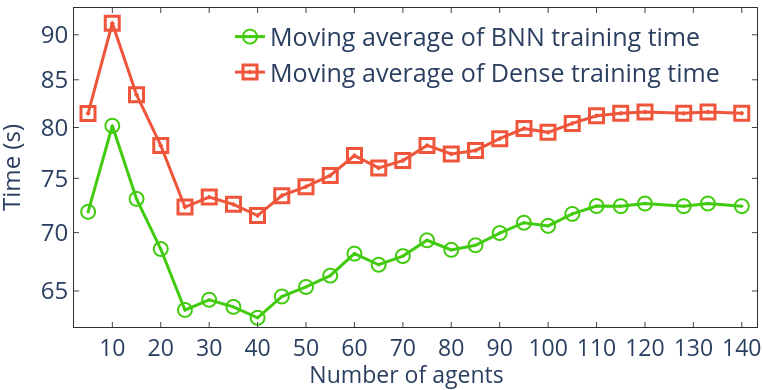}}
\renewcommand{\thesubfigure}{(f)}
\subfigure[]{\label{fig:f}\includegraphics[width=58mm]{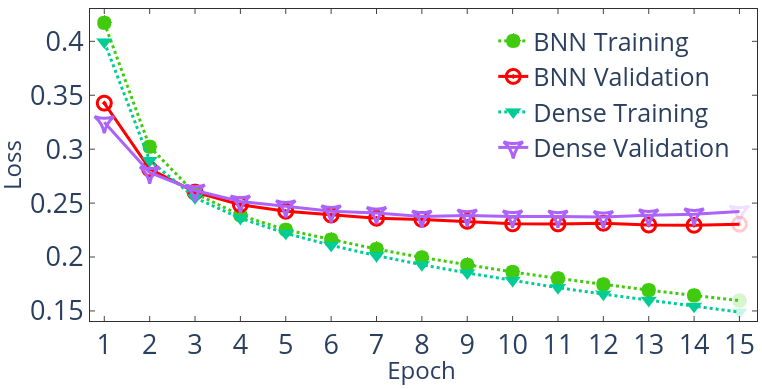}}
\caption{Experiment results: (a,b) ACC and RMSE of different approaches; (c,d) Moving average of prediction and training time for BNN compared to Dense; (e,f) Convergence of validation accuracy and loss for BNN compared to Dense.}
\end{figure*}

In Algorithm \ref{alg1}, the training data - which consists of $Features$ as given by Eq. \ref{deq:6} and $Label$ - is first initialized to $\emptyset$. Then, the first-hand evidence $E_a$ is being iterated over (line 3) to find out historical information about the agent $z$, i.e., the outcome $t$ and context $\varsigma_1,...,\varsigma_k$ of each interaction.  This information is then supplied to $G_a(.)$ (Eq. \ref{deq:6}) to obtain the predicted conditional probability $P\left(t=1 \left\vert  \varsigma_1,...,\varsigma_k\right.\right)$. The probabilities and the corresponding labels are then added to $Features$ to form the training data (lines 8-12).

After initialization, all the subsequent updates are performed using Algorithm \ref{alg2} and \ref{alg3}, where Algorithm \ref{alg2} is executed when a new first-hand evidence is available at $a$ and Algorithm \ref{alg3} is executed when $a$ receives a new model from a new advisor agent or an updated model from an existing advisor agent.

\begin{proposition}
The time complexity of Algorithm \ref{alg1} is $O(|E_a|\times|X_a|)$.
\end{proposition}

\begin{proposition}
The time complexity of Algorithm \ref{alg2} and \ref{alg3} is $O(|X_a|)$ and $O(|E_a|)$, respectively.
\end{proposition}

The training and retraining of the neural network using the above training dataset can be either performed by the agent itself or outsourced to fog computing \cite{yousefpour2019all}. Similarly is the storage of the neural network.

\section{Evaluation} \label{evSec}
We evaluate COBRA using both experiments and simulations. 

\subsection{Experiment setup} \label{ev1Sec}

{\bf Dataset.}
We use a public dataset obtained from \cite{Zheng:2014:IQR:2587728.2587740} which contains the response-time values of $4,532$ web services invoked by $142$ service users over $64$ time slices. The dataset contains $30,287,611$ records of data in total, which translates to a data sparsity of $26.5\%$. Following \cite{nielsen1994usability}, we assume a standard SLA which specifies that 1 second is the limit that keeps a user's flow of thought uninterrupted. Hence, response time above $1$ second is considered violation of SLA and assigned a \textit{False} label, while response time below or equal to $1$ second is assigned a \textit{True} label which indicates that the SLA is met. 

{\bf Platform.}
All measurements are conducted using the same Linux workstation with 12 CPU cores and 32GB of RAM. The functional API of {\tt Keras} 
is used for the implementation of the neural network architectures on top of {\tt TensorFlow} 
backend while {\tt scikit-learn} 
is used for the implementation of Gaussian process, decision tree, and Gaussian Naive Bayes models.

{\bf Benchmark methods.}
We use the following benchmarks for comparison:
\begin{itemize}
    \item \textit{Trust and Reputation using Hierarchical Bayesian Modelling (HABIT) }: This probabilistic trust model is proposed by \cite{habit} and uses Bayesian modelling in a hierarchical fashion to infer the probability of trustworthiness based on direct and third-party information and outperforms other existing probabilistic trust models.
    \item \textit{Trust Management in Social IoT (TMSIoT)}: This model is proposed by \cite{SIOT}, in which the trustworthiness of a service provider is a weighted sum of a node's own experience and the opinions of other nodes that have interacted with the service provider.
    \item \textit{Beta Reputation System (BRS)}: This well-known model as proposed by \cite{josang2002beta} uses the beta probability density function to combine feedback from various agents to calculate a trust score.
\end{itemize}

{\bf Evaluation metrics.}
We employ two commonly used metrics. One is  the accuracy defined as
\[ACC=\frac{{TP} + {TN}}{{TP} + {TN} + {FP} + {FN}}\]
where \textit{TP = True Positive}, \textit{FP = False Positive}, \textit{TN = True Negative}, and \textit{FN = False Negative}.
The other metric is the root mean squared error (RMSE) defined by
\[ RMSE(T,\hat{T}) = \sqrt{\frac{1}{m}\Sigma_{i=1}^{m}{(T_i -\hat{T}_i)^2}}\]
Where $T$ is the ground-truth trustworthiness and $\hat{T}$ is the predicted probability of trustworthiness and $m$ is the total number of predictions.

\subsection{Experiment procedure and results}

We run COBRA for each of the 142 web-service clients to predict whether a web-service provider $z$ can be trusted to meet SLA, given a context $\zeta$ which is the time slice during which the service was consumed. We experiment on $800,000$ random samples of the dataset due to two main considerations: (1) COBRA is a multi-agent approach but in the experiment we build all the models and BNNs on one machine, (2) the significantly high time and space complexity of the Gaussian process used in HABIT restricts us to work with a sample of the dataset. We employ $10$-fold cross validation and compare the performance of COBRA with the benchmark methods described in Section \ref{ev1Sec}. In COBRA-DT, decision tree is used for model encapsulation for all 142 agents, in COBRA-GNB, Gaussian Naive Bayes is used for the encapsulation for all 142 agents, and in a hybrid approach, COBRA-Hyb, decision tree is used for 71 randomly selected agents while Gaussian Naive Bayes is used for the rest. In HABIT the reputation model is instantiated using Gaussian process with a combination of \textit{dot product + white kernel} co-variance functions. In COBRA-DT/GNB/Hyb-B, our proposed neural network architecture in Section \ref{deep} (BNN) is used, while in COBRA-DT/GNB/Hyb-D, a fully connected feed-forward architecture (Dense) is used instead.

The results, as illustrated in Fig. \ref{fig:a} and Fig. \ref{fig:b}, indicate that all the versions of COBRA with Bernoulli neural engine outperform the benchmark methods, while without our proposed Bernoulli neural architecture, HABIT is competent to Dense version of COBRA-GNB. The choice of the encapsulation model only slightly affects the performance in hybrid mode, which suggests that the performance of COBRA is stable. 

Furthermore, we present the moving average of prediction and training time for BNN versions of COBRA compared to Dense versions of COBRA respectively in Fig. \ref{fig:c} and Fig. \ref{fig:d}. The results indicate that our proposed BNN architecture significantly reduces the time required for training and making predictions.

Moreover, as illustrated in Fig. \ref{fig:e}, the divergence between training accuracy and validation accuracy of BNN is significantly smaller than that of Dense. Similarly, Fig. \ref{fig:f} depicts a smaller divergence between training loss and validation loss of BNN compared to that of Dense. These results indicate that Dense is more prone to overfitting as the epochs increase. 

\begin{figure}[t]
\centering
\subfigure[]{\label{fig:3d}\includegraphics[width=0.65\linewidth]{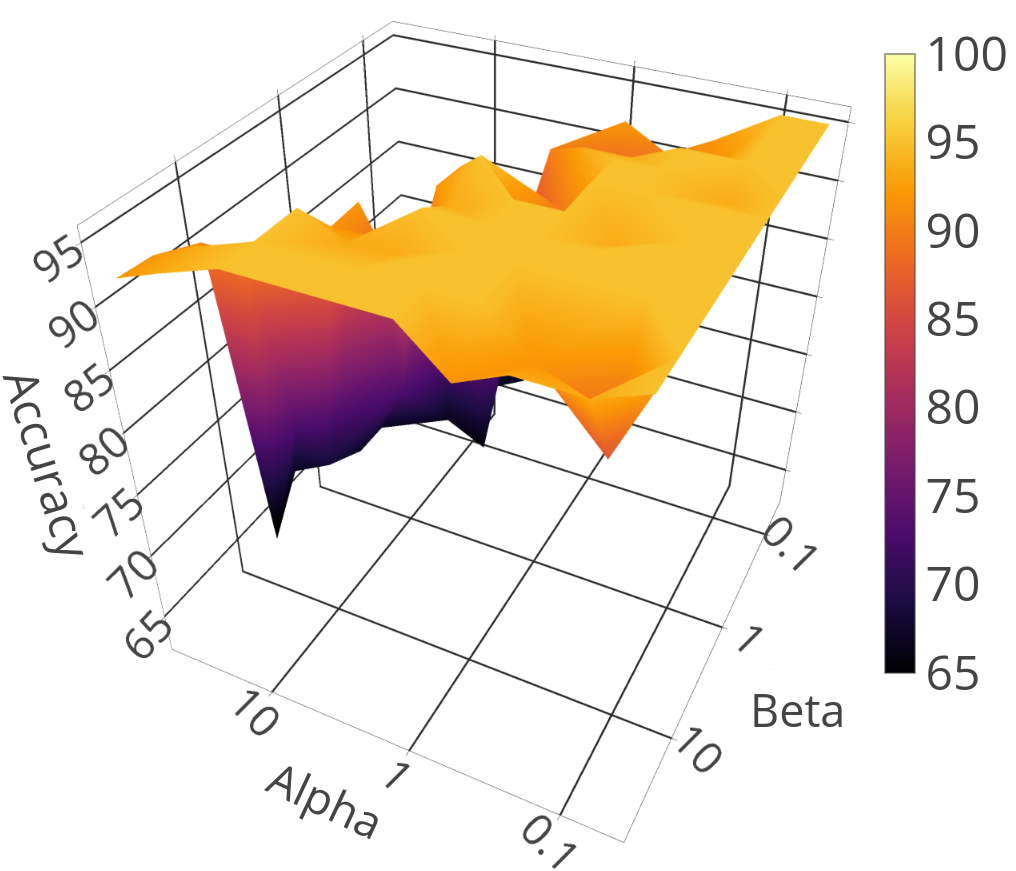}}
\subfigure[]{\label{fig:contour}\includegraphics[width=0.65\linewidth]{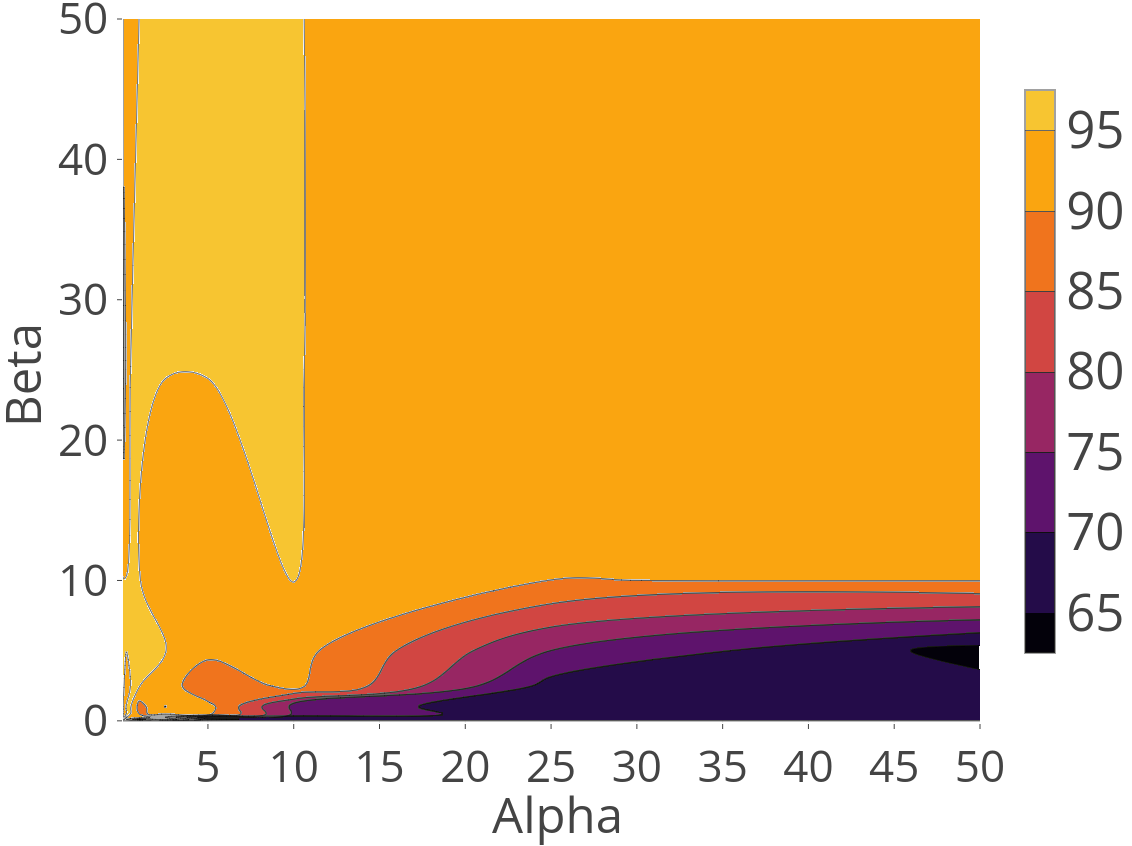}}
\caption{Simulation results: (a) Accuracy of COBRA (in percentage) versus different distributions of $\Phi$ parameterized by $\alpha$ and $\beta$ (in log scale). (b) Contour plot of (a); The color spectrum from {\it dark} to {\it light} represents accuracy from \textit{low} to \textit{high}.}
\end{figure}

\subsection{Simulation setup} \label{ev2Sec}
For a more extensive evaluation of COBRA, especially with respect to extreme scenarios which may not be observed often in the real world, we also conduct simulations. 

We simulate a multi-agent system with 51 malicious agents and 49 legitimate agents, in consideration of the 51 percent attack. The attack model used for malicious agents consists of fake and misleading testimonies which is a common attack in TRM systems. Specifically, a model shared by a malicious agent provides opposite prediction of the trustworthiness of a target agent, i.e., it outputs $1-p$ when the model would predict $p$ if it were not malicious.

Denote by $\Phi$ the probability that an arbitrary agent interacts with an arbitrary target agent, which we treat as a random variable with beta distribution parameterized by $\alpha$ and $\beta$. We run $100$ simulations each with a different distribution of $\Phi$. For example,  $\alpha=\beta=0.5$ means that one group of agents interact with the target agent frequently while another group seldom interact with the target agent; $\alpha=\beta=2$ means that most of the agents have half chance to interact with the target agent; $\alpha=5,\beta=1$ means that most of the agents interact with the target agent frequently, while $\alpha=1$ and $\beta=3$ means that most of the agents seldom interact with the target agent.

We use 4 synthesized context features randomly distributed in the range $[-1,1]$, and generate $100$ different target agents that violates SLA with a probability following the normal distribution on condition of each context feature.

\subsection{Simulation results}

The simulation results are shown in Fig. \ref{fig:3d}-\ref{fig:contour}, where the key observations are:
\begin{itemize}
  \item COBRA is able to predict accurate trust scores (probability of being trustworthiness) for the majority of the cases. Particularly, in 90 out of 100 simulated distributions of $\Phi$ an accuracy greater than or equal to 85\% is achieved.
  \item It is crucial to note that these results are achieved when 51\% of the agents are malicious. This shows that COBRA is resistant to the $51$ percent attack.
\end{itemize}


\section{Conclusion} \label{fut}
This paper proposes COBRA, a context-aware trust assessment framework for large-scale online environments (e.g., MAS and IoT) without a trusted intermediary. The main issue it addresses is an accuracy-privacy dilemma. Specifically, COBRA uses model encapsulation to preserve privacy that could otherwise be exposed by second-hand evidence, and in the meantime to retain context information as well. It then uses our proposed Bernoulli neural network (BNN) to aggregate the encapsulated models and first-hand evidence to make an accurate prediction of the trustworthiness of a target agent. Our experiments and simulations demonstrate that COBRA achieves higher prediction accuracy than state-of-the-art TRM systems, and is robust to 51 percent attack in which the majority of agents are malicious. It is also shown that the proposed BNN trains much faster than a standard fully-connected feed-forward neural network, and is less prone to overfitting.

\section{ Acknowledgments}
This work is partially supported by the MOE AcRF Tier 1 funding (M4011894.020) awarded to Dr. Jie Zhang.

\bibliographystyle{aaai}  
\bibliography{aaai2020}  

\end{document}